\documentclass[cameraready]{Interspeech}

\title{Disentangling Acoustic Cues in Alzheimer’s Pathology and Perception: The Roles of Language and Gender}

\author[affiliation={1} ]{Liu}{He}
\author[affiliation={2},correspondingauthor ]{Yuanchao}{Li}
\author[affiliation={1} ]{Yin-Long}{Liu}
\author[affiliation={1} ]{Rui}{Feng}
\author[affiliation={1} ]{Yiming}{Wang}
\author[affiliation={1} ]{Jiaxin}{Chen}
\author[affiliation={1} ]{Yizhe}{Wang}
\author[affiliation={1},correspondingauthor]{Jiahong}{Yuan}

\address{
    $^1$ University of Science and Technology of China\\
    $^2$ University of Edinburgh
}

\email{heliummn@mail.ustc.edu.cn, yuanchao.li@ed.ac.uk, jiahongyuan@ustc.edu.cn}


\newcommand{\blue}[1]{\textcolor{blue}{#1}}

\usepackage{comment}
\usepackage{multirow}
\usepackage{threeparttable}

\begin{document}

\maketitle

\begin{abstract}
Acoustic biomarkers show promise for detecting Alzheimer's Disease (AD), yet whether the cues driving diagnostic AI align with those salient to human listeners is underexplored across languages and genders, where pathological markers and perceptual strategies differ. We train models to predict clinical AD status (pathology) and human perceptual scores across Mandarin and Greek, male and female speakers. Using SHAP for interpretability and statistical models for validation, we compare feature importance by subgroup. Results reveal a context-dependent divergence: pathological–perceptual alignment is significant for Mandarin and female speakers but disappears for Greek and male speakers, where pathology models did not exceed chance—a failure mode population-specific auditing surfaces. Global Explainable AI (XAI) explanations can mask critical demographic divergences, highlighting the need for population-specific explainability auditing for equitable deployment of clinical speech AI.

\end{abstract}

\section{Introduction}

The analysis of spontaneous speech is a promising, non-invasive approach for detecting Alzheimer's Disease 
(AD), as the disease impacts the cognitive and fine motor 
control systems essential for speech production 
\cite{de2018changes, meilan2014speech}. A rich body of 
literature has identified robust acoustic markers across 
four key domains: temporal/fluency (e.g., increased pausing), prosodic (e.g., vocal monotony), phonatory 
(e.g., jitter, shimmer), and articulatory (e.g., reduced vowel space) \cite{meilan2014speech, parlak2023voice, 
shamei2023reduction}. Despite these advances, the clinical translation of speech-based AD detection faces a critical challenge: diagnostic models often operate as black boxes, limiting 
clinical trust. For these models to be adopted, their 
internal logic must be interpretable and, crucially, align with the acoustic cues that human listeners rely on—cues that
are perceptually accessible even across unfamiliar languages \cite{lai2007perception, verkhodanova2022expertise,10889008} and that shape the intelligibility of pathological speech
\cite{feenaughty2014relationship, jiang2023comprehension}. However, achieving this pathology-perception alignment is 
complicated by demographic diversity. Pathological biomarkers themselves vary
substantially across languages \cite{luz2024overview, 
perez2022alzheimer} and genders \cite{parlak2023voice, 
arslan2025effect}, and human
auditory judgments are likewise population-dependent, as cross-lingual
perception studies attest \cite{lai2007perception, verkhodanova2022expertise}. This heterogeneity means that a single, population-invariant alignment cannot be assumed: doing so risks deploying models that rely on cues imperceptible or misleading to human stakeholders for specific subgroups. While recent evaluations show that XAI 
reliability can vary across speaker groups \cite{wu2024can}, 
they focus on model-internal correctness rather than the 
external validity of explanations---specifically, whether 
XAI-highlighted cues align with what human stakeholders 
actually rely on. To our knowledge, no prior work has used XAI as an 
auditing framework to systematically compare the feature 
salience of pathology versus perception models across 
diverse demographic scenarios. Addressing this gap is a 
prerequisite for accountable deployment of speech AI in 
regulated clinical settings. To bridge this gap, this study trains models for two 
tasks: clinical AD classification (pathology) and 
prediction of human listener scores (perception), for 
Mandarin and Greek, male and female speakers. Using SHAP 
\cite{lundberg2017unified} for interpretability and GLMER 
\cite{bates2015fitting} for statistical validation, we 
directly contrast the feature importance profiles of these 
two tasks across language and gender subgroups.

\vspace{-5pt}

\section{Methods}

\subsection{Datasets and Participants}

In this study, we utilized Greek speech data from the ADReSS-M challenge and Mandarin speech samples from the NCMMSC2021 challenge to explore the potential impact of gender, language, and speech features on the perception of AD and HC groups \cite{luz2024overview}. All speech samples were elicited using a standardized picture description paradigm, ensuring consistency in speech task across all speakers and languages, thereby eliminating task type as a potential confound. For each language, 30 picture description utterances (15 AD/15 HC) with durations ranging from 0 to 1 minute were selected. Samples were selected to maximize balance across gender and diagnostic labels; complete parity was difficult due to limitations in the Greek corpus and the need to exclude samples with audio artefacts or incomplete metadata. This sample size is consistent with established practices in perceptual studies of pathological speech \cite{verkhodanova2022expertise}. 
Each utterance consisted of dialogues between healthcare professionals and participants. The audio characteristics of the Greek and Chinese recordings were generally comparable. The experiment included 30 audio recordings per language, with AD and HC samples randomly ordered within each language group. 
Sixteen students (8 male, 8 female) from the University of Science and Technology of China participated \cite{10.1007/978-981-95-5382-2_17}, with a mean age of 23 years, consistent with listener pool sizes in comparable crosslinguistic perception studies \cite{verkhodanova2022expertise,lai2007perception,10889008}. All participants were native Mandarin speakers without formal Greek training. The use of naive cross-lingual listeners was a deliberate design choice: prior research has established that listeners can reliably detect pathology- and disfluency-related acoustic cues in languages entirely unknown to them, demonstrating that such signals carry crosslinguistic perceptual validity independent of linguistic comprehension \cite{verkhodanova2022expertise,lai2007perception,10889008}. Variances in age and education levels between AD and HC groups across the two corpora motivated their inclusion as covariates in all subsequent statistical models. 
The stimuli were presented through the webMUSHRA interface \cite{schoeffler2018webmushra}. For each language condition, participants first received background information about AD and HC characteristics. They then listened to audio clips and performed two tasks: 1) classifying each sample as AD or HC, 2) identifying specific time segments influencing their decisions through a timeline annotation interface. Participants additionally provided free-text rationales for each classification. The Greek perception experiment preceded the Mandarin condition to prevent linguistic interference.




\subsection{Acoustic Feature Extraction}

All audio recordings were resampled to 16~kHz mono, from which we extracted 21 global features spanning four categories known to be affected by AD. \textbf{Temporal/Fluency} features were derived using a pre-trained Silero VAD (threshold=0.5, min\_speech=250~ms, min\_silence=100~ms), yielding \textit{Speech Portion}, pause metrics (\textit{number of pauses, total/mean pause duration, pause rate}), and \textit{Speech/Articulation Rate} computed from Whisper large-v3 transcripts~\cite{radford2022whisper}. \textbf{Prosodic} features were extracted via Parselmouth~\cite{parselmouth} using Praat's autocorrelation method with gender-specific pitch ranges (75--300~Hz male, 100--500~Hz female), yielding \textit{mean F0, F0 standard deviation, F0 range, mean F0 velocity}, and \textit{mean absolute F0 acceleration}. \textbf{Phonatory} features included \textit{Jitter}, \textit{Shimmer}, and \textit{HNR}. \textbf{Articulatory} features comprised \textit{mean F1} and \textit{F2} via Praat's Burg algorithm. For each stimulus, a \textbf{Perception Weighted Score (PWS)} was computed as the mean binary listener response (1=AD, 0=HC) across all 16 listeners.

\subsection{Modeling and Analysis Framework}

Our methodology consists of two machine learning tasks: 1) a pathology classification task to differentiate between speakers with AD and HC, and 2) a perception regression task targeting a PWS. After evaluating six algorithms (Logistic Regression, SVM, Random Forest, XGBoost, LightGBM, CatBoost), we selected Random Forest framework (RFClassifier and RFRegressor)  due to its superior and robust performance  across both tasks. Importantly, age and education are never used as inputs to the predictive models; they serve exclusively as covariates in the GLMER (described below). Gender defines the gender-specific subgroups rather than serving as a predictor. Final model performance was assessed via 5-fold cross-validation (StratifiedKFold for classification). We report AUC, F1-Score, and accuracy for classification, and PearsonR, RMSE, and MAE for regression. Statistical significance of each classification AUC (against chance, 0.5) and regression r (against 0) was assessed by a permutation test (5000 label permutations on the pooled cross-validation predictions). To interpret the models, we employed SHAP~\cite{lundberg2017unified, rudin2019stop} to compute global feature importance. This analysis was designed to compare the acoustic drivers of automated classification versus human perception, both on the full dataset and within language (Mandarin/Greek) and gender (Male/Female) subgroups. SHAP values were computed per fold and averaged across all five folds to ensure stability of feature importance rankings. To statistically validate the influence of these factors on human perception, we complemented the SHAP analysis with a Generalized Linear Mixed-Effects Model (GLMER) \cite{bates2015fitting}. The GLMER was used to predict listeners' binary judgments based on fixed effects for language, gender, and key acoustic biomarkers. Speaker age and education level were included as covariates to control for potential demographic confounds; age differed significantly between AD and HC groups in the Greek corpus ($t=2.53$, $p=0.017$), motivating their inclusion. Listener variability was modelled as a random effect; the random effect variance was negligible ($\approx 0$), indicating high consistency across listeners.

\begin{table}[t]
\centering
\caption{Random Forest performance of pathology classification and perception
regression models across subgroups. Significance of each AUC (vs.\ chance) and
$r$ (vs.\ 0) assessed by a permutation test.}
\vspace{-10pt}
\label{tab:model_performance}
\begin{threeparttable}
\begin{tabular}{@{}lcccccc@{}}
\toprule
\multirow{2}{*}{\textbf{Task}} & \multicolumn{3}{c}{\textbf{Pathology}}
& \multicolumn{3}{c}{\textbf{Perception}} \\
\cmidrule(lr){2-4}\cmidrule(lr){5-7}
& AUC\,$\uparrow$ & Acc.\,$\uparrow$ & F1\,$\uparrow$
& $r$\tnote{a}\,$\uparrow$ & RMSE\,$\downarrow$ & MAE\,$\downarrow$ \\
\midrule
All & 0.70\textsuperscript{*}  & 0.70 & 0.70 & 0.72\textsuperscript{*} & 0.25 & 0.19 \\
CN  & 0.83\textsuperscript{*}  & 0.83 & 0.83 & 0.84\textsuperscript{*} & 0.20 & 0.14 \\
GK  & 0.60\textsuperscript{ns} & 0.60 & 0.60 & 0.54\textsuperscript{*} & 0.28 & 0.24 \\
M   & 0.52\textsuperscript{ns} & 0.52 & 0.52 & 0.73\textsuperscript{*} & 0.25 & 0.18 \\
F   & 0.79\textsuperscript{*}  & 0.79 & 0.79 & 0.74\textsuperscript{*} & 0.23 & 0.19 \\
\bottomrule
\end{tabular}
\begin{tablenotes}[flushleft]\footnotesize
\item[a] Pearson's correlation coefficient ($r$).
\item CN=Mandarin/Chinese, GK=Greek, M=Male, F=Female.
\item[*] significantly above chance, permutation test ($5000$ perms, $p<0.05$);
\textsuperscript{ns} not significant.
\end{tablenotes}
\end{threeparttable}
\vspace{-20pt}
\end{table}

\section{Results}

\subsection{Model Performance}

The performance of the Random Forest models, summarized in Table \ref{tab:model_performance}, reveals significant variability across linguistic and gender subgroups. The \textbf{Mandarin} models substantially outperformed all others in both pathology classification (AUC=0.83) and perception regression (Pearson's $r=0.84$). Conversely, the \textbf{Greek} models yielded the most modest results. A notable interaction emerged for gender: in the classification task, the \textbf{female-specific} model was highly effective (AUC=0.79), while the \textbf{male-specific} model performed at chance level (AUC=0.52). This disparity vanished in the perception task, where both male ($r=0.73$) and female ($r=0.74$) models performed similarly well. This performance gap, particularly for the male models, motivates the subsequent interpretability analysis. A permutation test (5000 permutations) confirmed that the male and Greek pathology models did not significantly exceed chance ($p>0.05$), whereas all perception models and the remaining pathology models did ($p<0.05$).

\subsection{Pathological vs. Perceptual Biomarkers}

\begin{figure*}[htb]

\begin{minipage}[b]{1\linewidth}
  \centering
  \centerline{\includegraphics[width=17 cm]{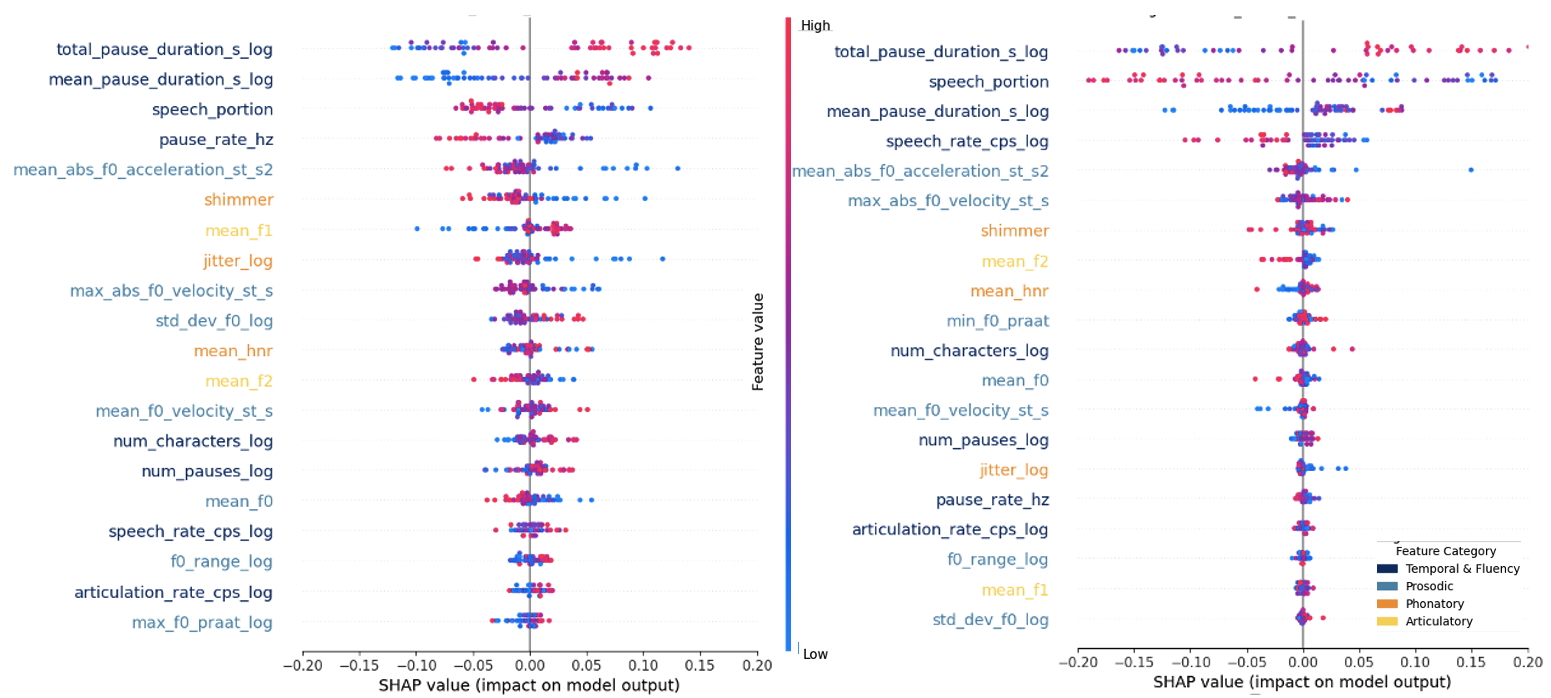}}
\end{minipage}
\vspace{-5pt}
\caption{Global SHAP feature importance across all data. \textbf{left} Pathology (classification), \textbf{right} Perception (human ratings)}
\label{fig:res}
\vspace{-15pt}
\end{figure*}

To identify the acoustic features driving each model's decisions, we analyzed and compared the SHAP feature importance rankings, as shown in Figure~\ref{fig:res}. Overall, a statistically significant, moderate positive correlation was found between the feature rankings of the pathology and perception models (Spearman's $\rho = 0.55, p < 0.01$). This indicates a partial, yet incomplete, alignment between the cues for automated diagnosis and those salient to human listeners. Despite this correlation, the models prioritize different acoustic domains. The pathology model relies on a combination of \textbf{Temporal/Fluency} features (e.g., \textit{pause\_rate\_hz}), \textbf{Articulatory} precision (e.g., \textit{mean\_f1}), and \textbf{Phonatory} instability (e.g., \textit{jitter\_log}). In contrast, the perception model is more sensitive to overall speech pace and activity, prioritizing different \textbf{Temporal/Fluency} metrics (e.g., \textit{speech\_rate\_cps\_log}, \textit{total\_pause\_duration\_s\_log}) and core \textbf{Prosodic} features (e.g., \textit{min\_f0\_praat}). This suggests that the diagnostic model identifies subtle, multi-domain disruptions, while the perception model captures more holistic, listener-centric changes in speech rhythm and pitch.

\subsection{Cross-Lingual and Cross-Gender Analysis}

The moderate alignment observed overall is largely driven by specific subgroups, as the relationship between pathological and perceptual biomarkers varies significantly across both language and gender. 

\noindent\textbf{Cross-Lingual Analysis.} A sharp divergence was found between languages. For Mandarin speakers, the alignment between pathology and perception feature rankings remained significant ($\rho = 0.54, p < 0.05$). For Greek speakers, however, this correlation vanished completely ($\rho = 0.10, p > 0.05$). This is because the underlying feature importance profiles are fundamentally different for each language (Pathology: Chinese vs. Greek, $\rho \approx -0.07$; Perception: Chinese vs. Greek, $\rho \approx 0.01$). The Mandarin pathology model prioritizes \textbf{Temporal/Fluency}
features (\textit{mean\_pause\_duration\_s\_log} and
\textit{pause\_rate\_hz}), consistent with F0 being phonologically
constrained in a tonal language
\cite{duanmu2007phonology,yuan2011perception}. The Greek pathology
model instead showed its highest SHAP values for \textbf{Prosodic}
variability (\textit{std\_dev\_f0\_log}); however, as this model did
not significantly exceed chance (Table~\ref{tab:model_performance}),
this ranking should be read with caution and may reflect
sample-specific patterns or demographic confounds (e.g., age) rather
than a robust disease- or language-related effect.

\noindent\textbf{Cross-Gender Analysis.} The pathology–perception
alignment was significant for female speakers ($\rho = 0.52, p < 0.05$)
but non-existent for males ($\rho = 0.06, p > 0.05$). For females, both
models were reliable and consistently driven by \textbf{Temporal/Fluency}
features (particularly pausing behavior), so this alignment indicates
that diagnostically relevant cues are also perceptually salient. For
males, however, the pathology model did not significantly exceed chance;
its near-zero alignment with the well-performing perception model
therefore reflects this unreliability rather than a genuine divergence
in diagnostic versus perceptual cues---precisely the failure mode that
population-specific auditing is designed to surface.

\subsection{Statistical Modeling of Perceptual Factors}

To statistically validate our findings, we first constructed a GLMER to predict listeners' binary perception of AD. As shown in Table~\ref{tab:glmer_singlecol}, after controlling for listener variability, several acoustic features were highly significant predictors. Notably, longer pause duration (\textit{MeanPauseDur\_log\_z}) and greater F0 standard deviation (\textit{StdDevF0\_log\_z}) significantly increased the chance of an AD percept, while higher shimmer unexpectedly decreased it. After controlling for speaker age and education (both significant positive predictors, see in Table~\ref{tab:glmer_singlecol}), the effect of Language reached significance ($\beta = 0.846$, $p = 0.043$), suggesting Greek speech was more likely to be perceived as AD once demographic differences were accounted for. The independent effect of Gender remained non-significant.

\begin{table}[t]
\centering
\scriptsize
\setlength{\tabcolsep}{2.5pt}

\caption{GLMER results predicting listener perception (1=AD). Reference groups: Male, Chinese.}
\label{tab:glmer_singlecol}
\vspace{-10pt}
\resizebox{\columnwidth}{!}{%
\begin{tabular}{l
                r r l
                r r l}
\toprule
& \multicolumn{3}{c}{\textbf{Main Effects}} 
& \multicolumn{3}{c}{\textbf{Gender Interaction}} \\
\cmidrule(lr){2-4} \cmidrule(lr){5-7}
Predictor 
& $\beta$ & SE & $p$
& $\beta$ & SE & $p$ \\
\midrule
(Intercept) & -0.300 & 0.259 & 0.248 
            & -0.298 & 0.332 & 0.370 \\
L (GK) & 0.846$^{*}$ & 0.419 & 0.043 
       & 1.238$^{*}$ & 0.488 & 0.011 \\
G (F) & -0.151 & 0.161 & 0.347 
      & -0.380$^{.}$ & 0.194 & 0.050 \\
Mean Pause Dur. & 1.592$^{***}$ & 0.131 & $<$.001
                 & 2.194$^{***}$ & 0.287 & $<$.001 \\
$\times$ G (F) &  &  & 
                & -0.747$^{*}$ & 0.334 & 0.025 \\
Shimmer & -0.813$^{***}$ & 0.180 & $<$.001
         & -0.678$^{**}$ & 0.212 & 0.001 \\
$\times$ G (F) &  &  & 
                & -0.368$^{*}$ & 0.186 & 0.048 \\
Std. Dev. F0 & 0.607$^{***}$ & 0.128 & $<$.001
              & 0.227 & 0.197 & 0.249 \\
$\times$ G (F) &  &  & 
                & 0.581$^{*}$ & 0.233 & 0.013 \\
Mean F2 & -0.052 & 0.084 & 0.536
         & 0.780$^{***}$ & 0.153 & $<$.001 \\
$\times$ G (F) &  &  & 
                & -1.411$^{***}$ & 0.182 & $<$.001 \\
Age & 0.236$^{**}$ & 0.074 & 0.001
     & 0.197$^{*}$ & 0.078 & 0.011 \\
Education & 0.228$^{**}$ & 0.085 & 0.008
           & 0.311$^{***}$ & 0.094 & $<$.001 \\
\bottomrule
\end{tabular}%
}

\vspace{2pt}
{\scriptsize $^{.}p<0.1$;\ $^{*}p<0.05$;\ $^{**}p<0.01$;\ $^{***}p<0.001$.}
\vspace{-20pt}
\end{table}

Our machine learning results suggested that the salience of acoustic cues is modulated by gender. To formally test this, we introduced interaction terms between gender and the acoustic features into the GLMER. The results, shown in Table~\ref{tab:glmer_singlecol}, revealed highly significant 
interactions. The perceptual effect of F0 standard deviation 
(\textit{StdDevF0\_log\_z}) was significantly stronger for 
female speakers (Interaction: $\beta = 0.581$, $p = 0.013$). 
Most interestingly, the effect of the second formant 
(\textit{MeanF2\_log\_z}) reversed direction entirely, being 
a positive predictor for males but a negative predictor for 
females (Interaction: $\beta = -1.411$, $p < 0.001$). These 
findings statistically confirm that listeners employ different 
auditory strategies when evaluating male and female speech 
for signs of AD, validating the gender-specific patterns 
observed in our SHAP analysis.

\section{Discussion and Conclusion}
Our study reveals a critical divergence between the acoustic biomarkers relevant for automated AD diagnosis and those salient to human perception. While we found a moderate overall alignment, this relationship is highly context-dependent, breaking down under the influence of speaker's language and gender.

\noindent\textbf{Divergence in Pathological vs. Perceptual Cues.} The discrepancy between the pathology and perception models suggests they capture different aspects of speech degradation. The pathology model, trained on clinical ground truth, identifies subtle disruptions across multiple acoustic domains, including \textbf{phonatory} (\textit{jitter}) and \textbf{articulatory} (\textit{mean\_f1}) features that may be less obvious to the naked ear. In contrast, the perception model, which mimics human listeners, prioritizes more holistic and immediately apparent cues related to speech rhythm and pacing (\textbf{Temporal/Fluency}) and overall pitch (\textbf{Prosodic}). This is directly confirmed by our own GLMER results, where pause duration emerged as the single most powerful predictor of listener judgments, and F0 standard deviation was also highly significant. Interestingly, our statistical analysis uncovered two counter-intuitive perceptual patterns with direct clinical implications. First, higher shimmer was associated with a lower probability of an AD percept. Rather than interpreting phonatory roughness as a sign of cognitive decline, naive listeners may associate irregular voice quality with vocal effort or emotional expressiveness, perceptual heuristics more consistent with a healthy but fatigued voice than a pathological one \cite{eadie2006effect}. This suggests that shimmer, while diagnostically relevant for the automated model, is perceptually misleading and could cause clinicians relying on auditory impression to systematically underestimate AD severity in speakers with prominent voice roughness. Second, greater F0 standard deviation increased the likelihood of an AD percept. This is counter-intuitive because AD is classically associated with vocal monotony \cite{meilan2014speech,parlak2023voice}. A likely explanation is that naive listeners flag any atypical prosody as abnormal, rather than specifically monotony, a perceptual overgeneralization with important consequences for clinical training and human-in-the-loop diagnostic design.

\noindent\textbf{Cross-Lingual and Gender Effects.} Most notably, the context-dependency of this alignment is revealed through cross-lingual and cross-gender comparisons. The alignment holds for Mandarin speakers but disappears for Greek speakers, reflecting both linguistic differences and the limited reliability of the Greek pathology model. In Mandarin, a tonal language, F0 is phonologically constrained to encode lexical meaning \cite{duanmu2007phonology}, so AD-related decline is more likely to disrupt temporal and fluency dimensions that are less phonologically constrained \cite{de2018changes,meilan2014speech}. This explains why both the Mandarin pathology and perception models prioritize temporal features, consistent with evidence on the unique role of F0 in Mandarin speech perception \cite{yuan2011perception}. 
In contrast, Greek, a non-tonal language with a rich intonation system \cite{arvaniti2005intonational}, places fewer phonological constraints on F0, potentially allowing cognitive decline to manifest through prosodic variability \cite{arvaniti2007greek}. However, the Greek corpus exhibited a significant age imbalance ($t = 2.53$, $p = 0.017$). Because the classifier received only acoustic
features as input (Section 2.3) and had no direct access to age, it
may nonetheless have partially captured age-correlated prosodic
variation through these features rather than purely disease-specific
cues. Moreover, as the Greek pathology model did not significantly exceed chance (Table 1), its SHAP profile—like that of the male model—should be interpreted with caution rather than as a stable clinical signature. Our naive Mandarin-speaking listeners, lacking the reference frame for Greek intonation norms, could not interpret these prosodic deviations as pathological \cite{lai2007perception}, likely explaining why the alignment collapses for Greek. This illustrates a broader deployment risk: without population-specific explainability auditing, diagnostic models may rely on demographic-confounded cues that misalign with stakeholder expectations. An even more powerful divergence was observed across genders. The alignment was significant for females, whose pathology and perception models were both consistently driven by pausing behavior. For males, this alignment disappeared entirely. Critically, the male pathology model performed at chance level (AUC=0.52), indicating it failed to learn a generalizable diagnostic signal from this limited sample. Its SHAP profile, dominated by \textbf{Phonatory} cues (\textit{shimmer}) rather than the temporal markers that proved robust for other subgroups, should therefore be interpreted with caution, as it may reflect sample-specific patterns rather than stable clinical signatures. By contrast, the male perception model performed well ($r = 0.73$), relying on fluency and \textbf{Articulatory} features (\textit{mean\_f2}) \cite{shamei2023reduction}. This asymmetry is itself informative: our XAI auditing framework reveals not only where pathology-perception alignment breaks down, but also when a model's internal logic has become unreliable for a specific demographic subgroup, a failure mode that would remain invisible without population-specific explainability analysis. This finding adds a technical dimension to growing evidence that gender is a critical variable in the speech AI ecosystem \cite{williams2025public}. That XAI explanations proved unreliable for specific subgroups echoes
recent evidence that local XAI explanations in speech are unstable
under distribution shift \cite{hjuler2025exploring}. While that work
examined emotional speech, our results extend this concern to a
clinical diagnostic context, reinforcing the need to evaluate XAI
reliability in applied speech tasks \cite{wu2024can}

\noindent\textbf{Limitations and Future Work.} We acknowledge several limitations in this exploratory
study. The sample size of 30 utterances per language, while consistent with comparable perceptual paradigms in pathological speech research \cite{verkhodanova2022expertise}, remains modest and may limit the stability of subgroup-level findings; we therefore frame our cross-lingual and cross-gender analyses as hypothesis-generating rather than confirmatory. This constraint reflects well-documented challenges in pathological speech data collection \cite{yue2025challenges}. The listener panel was culturally homogeneous; while naive cross-lingual listeners can detect pathology-related acoustic cues independently of linguistic comprehension \cite{verkhodanova2022expertise, lai2007perception}, their strategies may not generalize to trained clinicians or native speakers \cite{kreiman1990listener}. In particular, the Greek perception results should be interpreted as reflecting naive acoustic-driven judgments. Future work should expand to larger multilingual datasets, incorporate expert clinical listeners, and evaluate multiple XAI methods to assess explanation stability, contributing to demographic-aware evaluation frameworks for trustworthy speech AI deployment \cite{williams2025public}.

\section{Acknowledgments}
We thank the members of the Student Club of Languages and Linguistics
at the University of Science and Technology of China and other
volunteers for participating in the perception experiment, and
Yingchuan Yang for assistance with participant recruitment.
This work is supported in part by the National Social Science Foundation of China (Grant No. 23AYY012) and by the Super-computing Center of the University of Science and Technology of China. 

\section{Generative AI Use Disclosure}
Generative AI tools were used only for minor improvements in language and presentation. 
No AI system was used to generate, modify, or interpret the scientific content of this manuscript.
\vspace{-5pt}

\bibliographystyle{IEEEtran}
\bibliography{mybib}

\end{document}